\documentclass[prc,preprint]{revtex4}
\usepackage{graphics}

\begin{document}

\title{\bf Finite temperature calculations for the bulk properties of strange
star using a many-body approach}
\author{\bf G.H. Bordbar $^{1,2}$\footnote{Corresponding author}
\footnote{E-Mail: bordbar@physics.susc.ac.ir}, A. Poostforush $^1$
and A. Zamani $^1$}
  \affiliation{$^1$Department of Physics, Shiraz University,
Shiraz 71454, Iran\footnote{Permanent address},\\
and\\
$^2$Research Institute for Astronomy and Astrophysics of Maragha,\\
P.O. Box 55134-441, Maragha, Iran}

\begin{abstract}
We have considered a hot strange star matter, just after the
collapse of a supernova, as a composition of strange, up and down
quarks to calculate the bulk properties of this system at finite
temperature with the density dependent bag constant.
To parameterize the density dependent bag constant, we use our
results for the lowest order constrained variational (LOCV)
calculations of asymmetric nuclear matter.
 Our calculations for the structure properties of the
strange star at different temperatures indicate that its maximum
mass decreases by increasing the temperature.
We have also compared our results with those of a fixed value of
the bag constant.
It can be seen that the density dependent bag constant leads to
higher values of the maximum mass and radius for the strange star.
\end{abstract}
\maketitle


\section{Introduction}
\label{S:intro}
Strange stars are those which are built mainly from self bound
quark matter.
The surface density of strange star is equal to the density of
strange quark matter at zero pressure ($\sim 10^{15}\ g/cm^3$),
which is fourteen orders of magnitude greater than the surface
density of a normal neutron star. The central density of these
stars is about five times greater than the surface density
\cite{haensel,glendening,weber}.
The existence of strange stars which are made of strange quark
matter was first proposed by Itoh \cite{a} even before the full
developments of QCD.
Later Bodmer \cite{b} discussed the fate of an astronomical object
collapsing to such a state of matter.
In 1970s, after the formulation of QCD, the perturbative
calculations of the equation of state of the strange quark matter
was developed, but the region of validity of these calculations
was restricted to very high densities \cite{collins}.
The existence of strange stars was also discussed by Witten
\cite{c}.
He conjectured that a first order QCD phase transition in the
early universe could concentrate most of the quark excess in dense
quark nuggets. He suggested that the true state of matter was
strange quark matter.
Based on theoretical works of Witten on cosmic separation of
phases, the transition temperature is approximately $100\ MeV$, an
acceptable QCD temperature \cite{c}.
Witten proposal was that the strange quark matter composed of
light quarks is more stable than nuclei, therefore  strange quark
matter can be considered as the ground state of matter.
The strange quark matter would be the bulk quark matter phase
consisting of almost equal numbers of up, down and strange quarks
plus a small number of electrons to ensure the charge neutrality.
A typical electron fraction is less than $10^{-3}$ and it
decreases from the surface to the center of strange star
\cite{haensel,glendening,weber}.
Strange quark matter would have a lower charge to baryon ratio
compared to the nuclear matter and can show itself in the form of
strange stars \cite{c,d,e,f}.

Just after the collapse of a supernova, a hot strange star may be
formed.
A strange star may be also formed from a neutron star and is
denser than the neutron star.
If sufficient additional matter is added to a strange star, it
will collapse into a black hole. Neutron stars with masses of
$1.5-1.8 M_\odot$ with rapid spins are theoretically the best
candidates for conversion to the strange stars. An extrapolation
based on this indicates that up to two quark-novae occur in the
observable universe each day.
Besides, recent Chandra observations indicate that objects RX
J185635-3754 and 3C58 may be bare strange stars \cite{prakash}.

In this article, we consider a hot strange star born just after
the collapse of a supernova. Here we ignore the effects of the
presence of electrons, and consider a strange star purely made up
of the quark matter consisting of the up, down and strange quarks.
The energy of quark matter is calculated at finite temperature,
and then its equation of state is derived. Finally using the
equation of state of quark matter, the structure of strange star
at different temperatures is computed by integrating the
Tolman-Oppenheimer-Volkoff (TOV) equations.

\section{Calculation of Quark Matter Equation of State}\label{Form}
\subsection{Density Dependent Bag Constant}
Different models have been used for deriving the equation of state
of quark matter. Therefore there is a great variety of the
equations of state for this system. The model which we use is the
MIT bag model which was developed to take into account the non
perturbative effects of quark confinement by introducing the bag
constant. In this model, the energy per volume for the quark
matter is equal to the kinetic energy of the free quarks plus a
bag constant (${\cal B}$) \cite{chodos}. The bag constant ${\cal
B}$ can be interpreted as the difference between the energy
densities of the noninteracting quarks and the interacting ones.
Dynamically it acts as a pressure that keeps the quark gas in
constant density and potential. This constant is shown to have
different values which are $55$ and $90\ \frac{MeV}{fm^3}$ in the
initial MIT bag model.
Since the density of strange quark matter increases from surface
to the core of the strange star, it is more appropriate to use a
density dependent bag constant rather than a fixed bag constant.

According to the analysis of the experimental data obtained at
CERN, the quark-hadron transition takes place at about seven times
the normal nuclear matter energy density ($ 156\ MeVfm^{-3}$)
\cite{aa,g}. Recently, a density dependent form  has been also
considered for ${\cal B}$ \cite{adami,jin,blasch,burgio}.
The density dependence of ${\cal B}$ is highly model dependent. In
this article, the density dependence of ${\cal B}$ will be
parameterized, and we make the asymptotic  value of ${\cal B}$
approach a finite value ${\cal B}_{\infty }$ \cite{burgio},

\begin{equation}\label{eq1}
{\cal B}(n)={\cal B}_{\infty}+({\cal B}_0-{\cal
B}_{\infty})e^{-\gamma(n/n_0)^2}.
\end{equation}
The parameter $ {\cal B}_{0} = {\cal B}(n = 0)$ has constant value
which is assumed to be $ {\cal B}_{0} =400\ \frac{MeV}{fm^3}$ in
this work, and $\gamma$ is the numerical parameter which is
usually equal to $n_{0}\approx 0.17fm^{-3}$, the normal nuclear
matter density. ${\cal B}_{\infty }$ depends only on the free
parameter $ {\cal B}_{0}$.
We know that the value of the bag constant (${\cal B}$) should be
compatible with experimental data. The experimental results at
CERN-SPS confirms a proton fraction $x_{pt}=0.4$ (data is from
experiment on accelerated Pb nuclei) \cite{aa,burgio}.
Therefore, in order to evaluate ${\cal B}_{\infty }$, we use the
equation of state of the asymmetric nuclear matter. The
calculations regarding this can be found in the next section.

\subsection{Computation of ${\cal B}_{\infty
}$ using the asymmetric nuclear matter calculations}
We use the equation of state of the asymmetric nuclear matter to
calculate ${\cal B}_{\infty }$. For calculating the equation of
state of asymmetric nuclear matter, we employ the lowest order
constrained variational (LOCV) many-body method based on the
cluster expansion of the energy as follows \cite{b2, b3, b4, b5,
b6, b7, b8, b9, b10}.

The asymmetric nuclear matter is defined as a system consisting of
$Z$ protons ($pt$) and $N$ neutrons ($nt$) with the total number
density $n = n_{pt} + n_{nt}$ and proton fraction
$x_{pt}=\frac{n_{pt}}{n}$, where $n_{pt}$ and $n_{nt}$ are the
number densities of protons and neutrons, respectively.
For this system, we consider a trial wave function as follows,
\begin{equation}
\psi =F\phi,
\end{equation}
where $\phi$ is the slater determinant of the single-particle wave
functions and $F$ is the A-body correlation operator ($A=Z+N$)
which is taken to be
\begin{equation}
F={\cal S}\prod _{i>j}f(ij)
\end{equation}
and ${\cal S}$ is a symmetrizing operator.
For the asymmetric nuclear matter, the energy per nucleon up to
the two-body term in the cluster expansion is
\begin{equation}\label{eq3}
E([f])=\frac{1}{A}\frac{<\psi\mid H\mid\psi >}{<\psi\mid\psi>}
=E_1+E_2 \cdot
\end{equation}
The one-body energy, $E_1$, is
\begin{equation}
E_1=\sum_{i=1}^{2} \sum _{k_i} \frac{\hbar^{2}{k_i^2}}{2m},
\end{equation}
where labels $1$ and $2$ are used for proton and neutron
respectively, and $k_i$ is the momentum of particle $i$.
The two-body energy, $E_2$, is
\begin{equation}
E_2=\frac{1}{2A}\sum_{ij}<ij\mid{\cal V}(12)\mid ij-ji>,
\end{equation}
where
\begin{equation}
{\cal V} (12)=-\frac{\hbar^2}{2m}[f(12),[\nabla_{12}^2,f(12)]]+
f(12)V(12)f(12) \cdot
\end{equation}
In the above equation, $f(12)$ and $V(12)$ are the two-body
correlation and nucleon-nucleon potential, respectively. In our
calculations, we use $UV_{14}+TNI$ nucleon-nucleon potential
\cite{Lagaris}.
Now, we minimize the two-body energy with respect to the
variations in the correlation functions subject to the
normalization constraint. From the minimization of the two-body
energy, we obtain a set of differential equations. We can
calculate the correlation functions by numerically solving these
differential equations. Using these correlation functions, the
two-body energy is obtained and then we can compute the energy of
asymmetric nuclear matter.
 The procedure of these
calculations has been fully discussed in reference \cite{b3}.


As it was mentioned in the previous section, the experimental
results at CERN-SPS confirms a proton fraction $x_{pt}=0.4$
\cite{aa,burgio},
therefore to compute ${\cal B}_{\infty }$, we proceed in the
following manner:
\begin{itemize}
 \item Firstly, we use our results of the previous section for the asymmetric nuclear
matter characterized by a proton fraction $x_{pt}=0.4$.
 By assuming that the hadron-quark transition takes
place at the energy density equal to $1100MeVfm^{-3}$
\cite{aa,burgio}, we find that the baryonic density of the nuclear
matter is $n_B=0.98fm^{-3}$ (transition density). At densities
lower than this value the energy density of the quark matter is
higher than that of the nuclear matter. With increasing the
baryonic density these two energy densities become equal at the
transition density, and above this value the nuclear matter energy
density remains always higher.
\item
Secondly, we determine $B_{\infty }=8.99\ \frac{MeV}{fm^3}$ by
putting the energy density of the quark matter and that of the
nuclear matter equal to each other.
\end{itemize}

\subsection{Calculations for the energy of quark matter at finite temperature}
To calculate the energy of quark matter, we need to know the
density of quarks in terms of the baryonic density. We do this by
considering two conditions of beta equilibrium and charge
neutrality. This leads to the following relations
\begin{equation}
    \mu_d=\mu_u-\mu_e,
\end{equation}
\begin{equation}
    \mu_s=\mu_u-\mu_e,
\end{equation}
\begin{equation}\label{eq2}
    \mu_s=\mu_d,
\end{equation}
\begin{equation}\label{eq3}
    2/3n_u-1/3n_s-1/3n_d-n_e=0,
\end{equation}
where $\mu_i$ and $n_i$ are the chemical potential and the number
density of particle $i$, respectively. As mentioned, we consider
the system as pure quark matter ($n_e=0$ ) \cite{d,n,o,p}. Thus
according to relation (\ref{eq3}), we have
\begin{equation}\label{eq4}
    n_u=1/2(n_s+n_d).
\end{equation}
The chemical potential, $\mu_i$, at any adopted values of the
temperature ($T$) and the number density ($n_i$) is determined by
applying the following constraint,
\begin{equation}\label{eq5}
    n_i=\frac{g}{2\pi^2}\int_0^{\infty}
    {f(n_i, k, T)}{k^2dk},
\end{equation}
where
\begin{equation}\label{eq9}
f(n_i, k,
T)=\frac{1}{Exp\{\beta((m_i^2c^4+\hbar^2k^2c^2)^{1/2}-\mu_i)\}+1}
\end{equation}
is the Fermi-Dirac distribution function \cite{qq}. In the above
equation, $\beta = \frac{1}{k_BT}$ and $g$ is the degeneracy
number of the system.

As it is previously mentioned, we consider the total energy of the
quark matter as the sum of the kinetic energy of the free quarks
and the bag constant (${\cal B}$). Therefore, the total energy per
volume of the quark matter (${\cal E}_{tot}$) can be obtained
using the following relation,
\begin{equation}\label{eq6}
    {\cal E}_{tot} = {\cal E}_u + {\cal E}_d + {\cal E}_s + {\cal B},
\end{equation}
where ${\cal E}_i$ is the kinetic energy per volume of particle
$i$,
\begin{equation}\label{eq7}
{\cal
E}_i=\frac{g}{2\pi^2}\int_0^{\infty}{(m_i^2c^4+\hbar^2k^2c^2)^{1/2}}
{f(n_i, k, T)}{k^2dk}.
\end{equation}

After calculating the energy, we can determine the other
thermodynamic properties of the system. The entropy of the quark
matter (${\cal S}_{tot}$) can be derived as follows
\begin{equation}\label{eq6}
    {\cal S}_{tot}={\cal S}_u + {\cal S}_d + {\cal S}_s,
\end{equation}
where ${\cal S}_i$ is the entropy of particle $i$,
\begin{eqnarray}\label{eq8}
{\cal S}_i(n_i, T)&=&-\frac{3}{\pi^2}k_B\int_0^{\infty} [f(n_i, k,
T)\ln(f(n_i, k,T))
    \nonumber\\&&
    +(1-f(n_i, k, T))\ln(1-f(n_i, k, T))]k^2dk.
\end{eqnarray}
The Helmholtz free energy per volume (${\cal F}$) is given by
\begin{equation}\label{eq10}
    {\cal F} = {\cal E}_{tot} - T{\cal S}_{tot}.
\end{equation}

\begin{figure}
\includegraphics{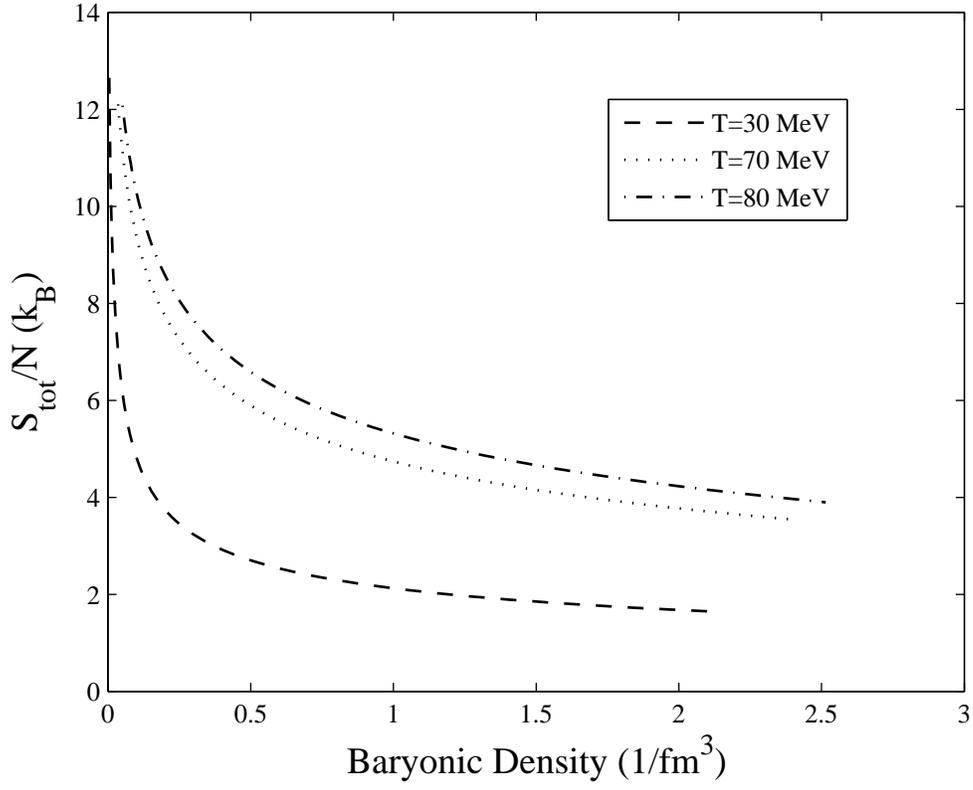}
\caption{The entropy per particle of the quark matter versus the
baryonic density at different temperatures for ${\cal B}=90\
\frac{MeV}{fm^3}$.} \label{fig1}
\end{figure}

\begin{figure}
\includegraphics{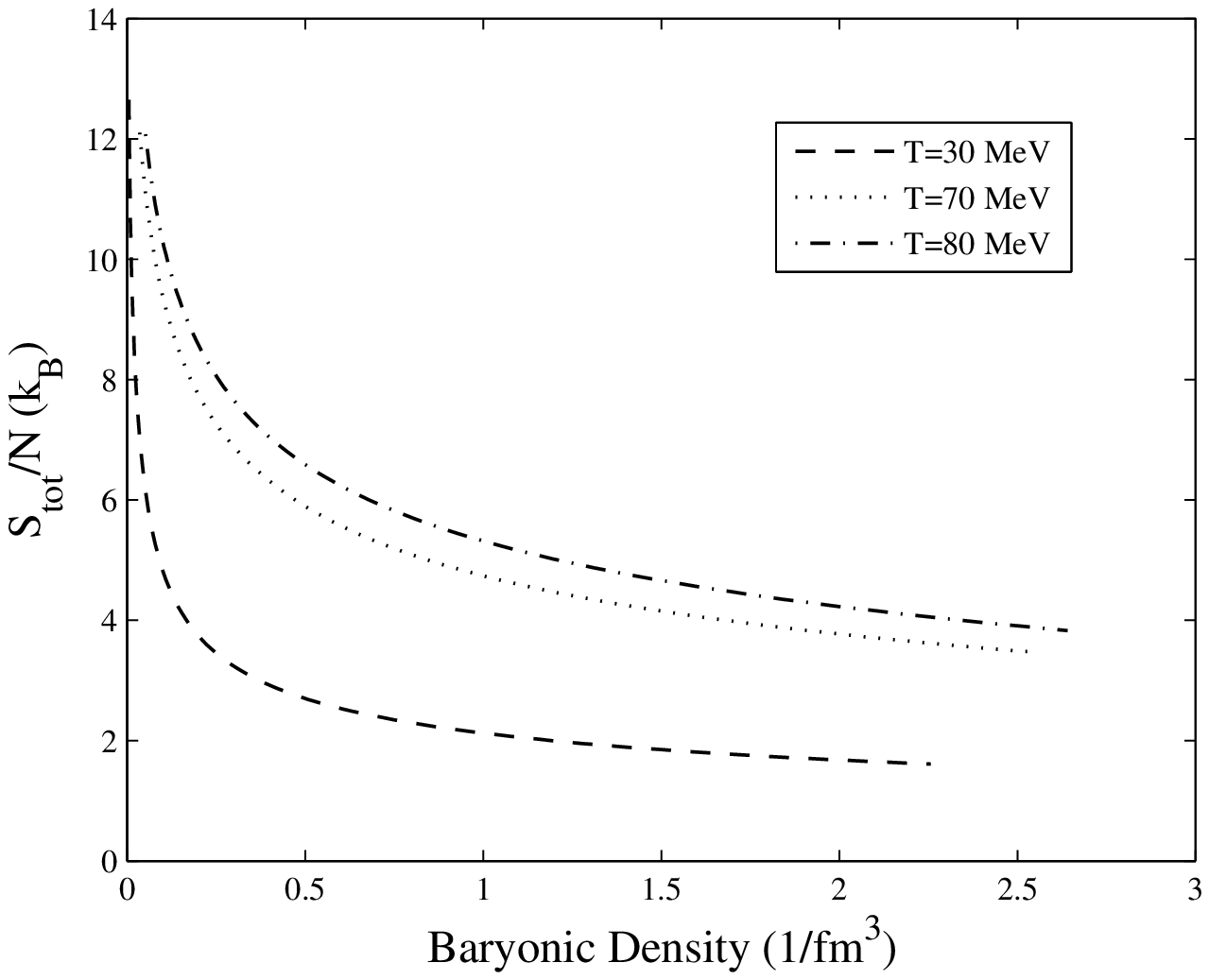}
\caption{As Figure \ref{fig1} but for the density dependent ${\cal
B}$.} \label{fig2}
\end{figure}

 The entropy per particle of the quark matter as a function of the
baryonic density for two cases of the constant and density
dependent ${\cal B}$ at different temperatures are plotted in
Figs. \ref{fig1} and \ref{fig2}. For a fixed temperature, we see
that the entropy per particle decreases by increasing the baryonic
density and for all relevant densities, it is seen that the
entropy increases by increasing the temperature.

\begin{figure}
\includegraphics{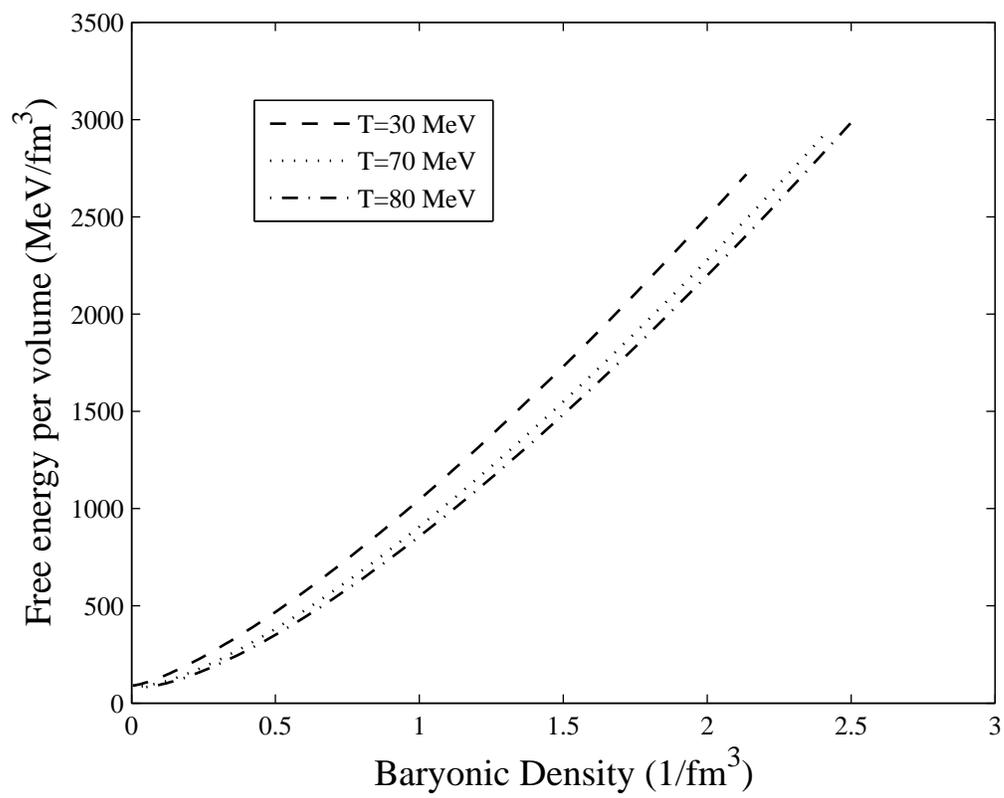}
\caption{The free energy per volume of the quark matter versus the
baryonic density  at different temperatures for ${\cal B}=90\
\frac{MeV}{fm^3}$.} \label{fig3}
\end{figure}

\begin{figure}
\includegraphics{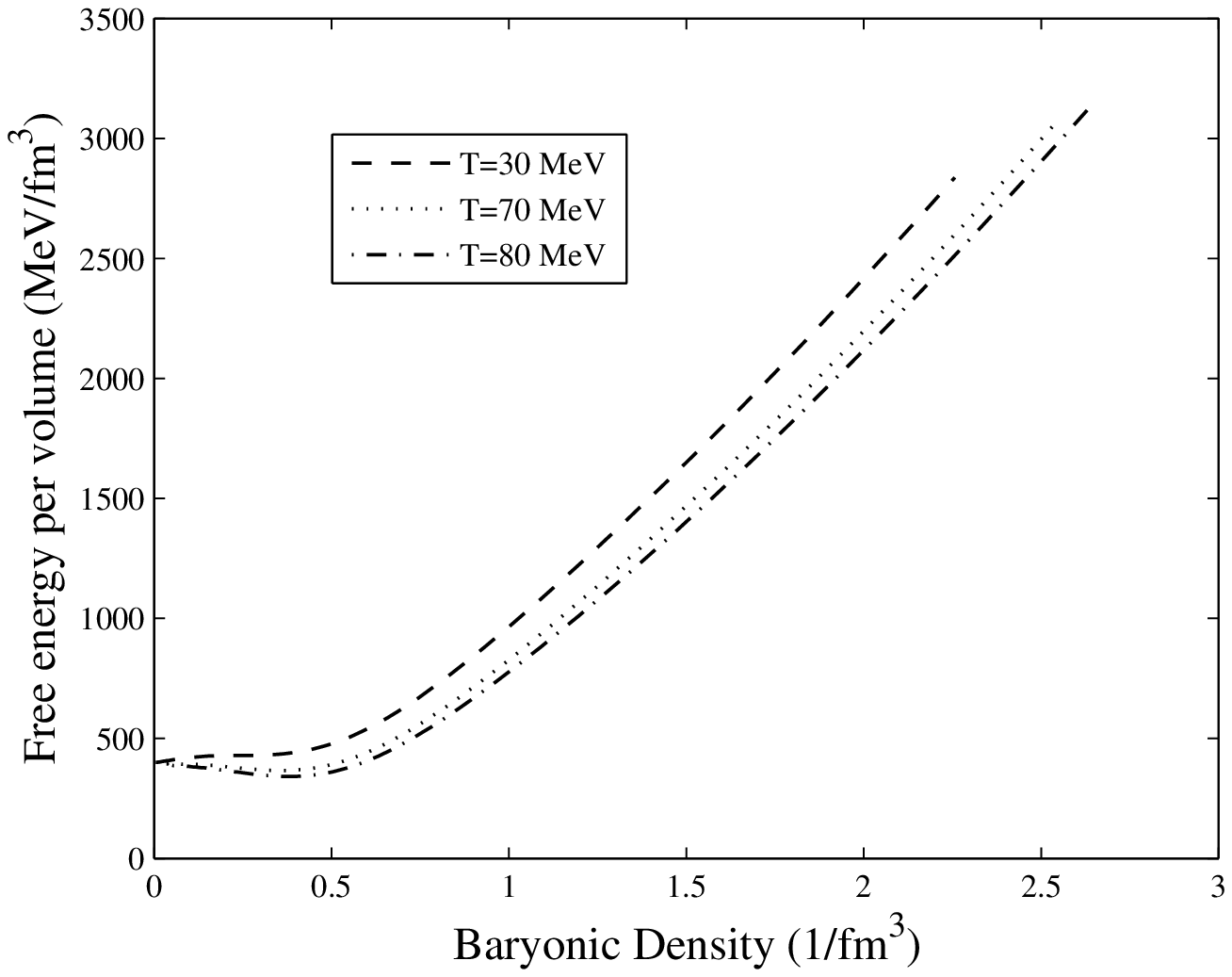}
\caption{As Figure \ref{fig3} but for the density dependent ${\cal
B}$.} \label{fig4}
\end{figure}

 In Figs. \ref{fig3} and \ref{fig4},
the free energy per volume of the quark matter versus the baryonic
density for two cases of the constant and density dependent ${\cal
B}$ are presented at different temperatures. We can see that the
free energy of the quark matter has positive values for all
densities and temperatures. For all densities, it is seen that the
free energy decreases by increasing the temperature.

\begin{figure}
\includegraphics{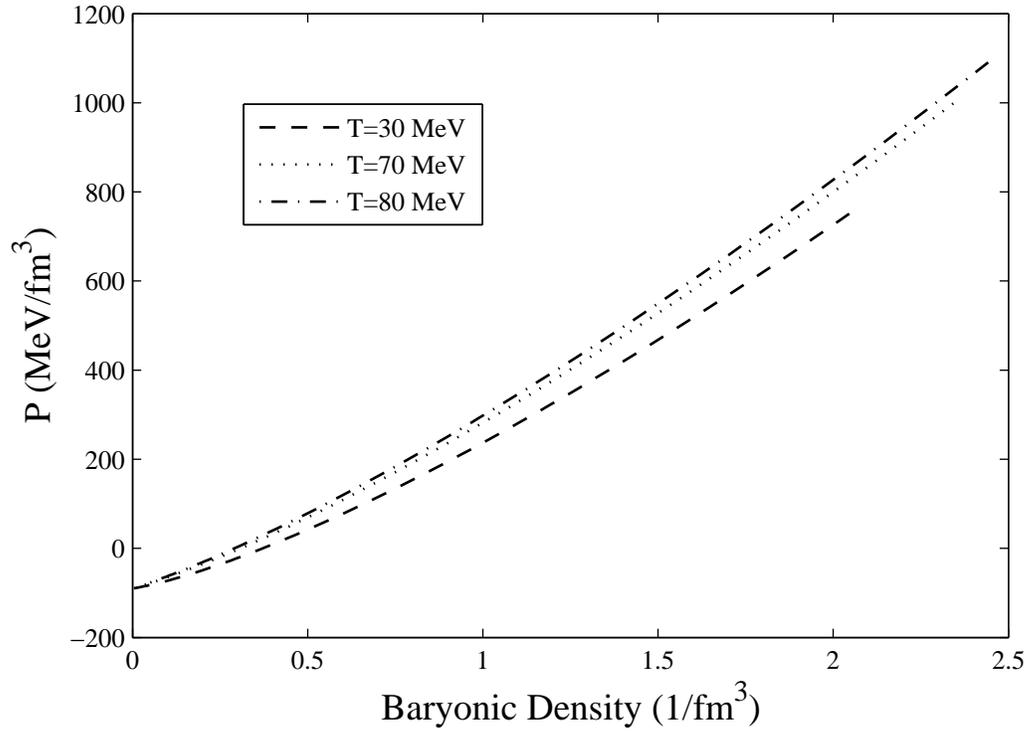}
\caption{The pressure of the quark matter as a function of the
baryonic density at different temperatures for ${\cal B}=90\
\frac{MeV}{fm^3}$. } \label{fig5}
\end{figure}

\begin{figure}
\includegraphics{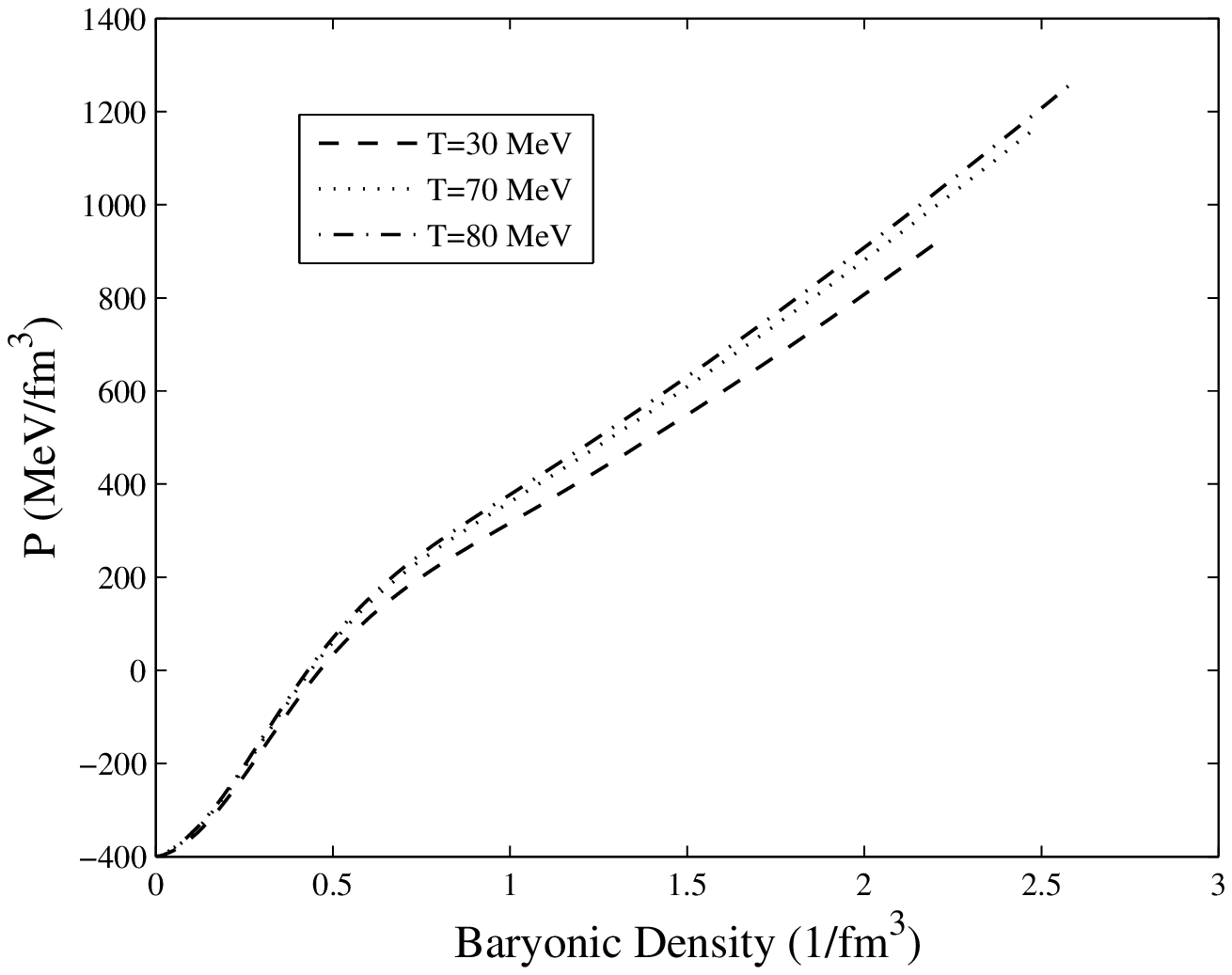}
\caption{As Figure \ref{fig5} but for the density dependent ${\cal
B}$.} \label{fig6}
\end{figure}

To obtain the structure of the strange star, the equation of state
of the quark matter is needed. For deriving the equation of state,
the following equation is used,
\begin{equation}\label{eq11}
P(n,T)=\sum_i {n_i \frac{\partial {\cal F}_i}{\partial n_i}-{\cal
F}_i},
\end{equation}
where $P$ is the pressure. The pressure of the quark matter versus
the baryonic density for two cases of the constant and density
dependent ${\cal B}$ are plotted in Figs. \ref{fig5} and
\ref{fig6}. It is seen that by increasing both density and
temperature, the pressure increases.
These figures show that for each temperature, the pressure becomes
zero at a specific value of the density.
We see that the density corresponding to zero pressure increases
by decreasing the temperature.

\section{Structure of Strange Star\label{Str}}
Compact objects like white dwarfs, neutron stars and strange stars
have limiting masses (maximum mass) and with a mass more than the
limitting value, the hydrostatic stability of the star is
impossible.
 For obtaining the maximum mass of the strange star, we
use the Tolman-Oppenheimer-Volkoff (TOV) equations \cite{n},

\begin{equation}\label{eq12}
    \frac{dP}{dr}=-\frac{G[{\cal E}(r)+\frac{P(r)}{c^2}]
    [m(r)+\frac{4\pi r^3 P(r)}{c^2}]}{r^2
    [1-\frac{2Gm(r)}{rc^2}]},
\end{equation}

\begin{equation}\label{eq13}
    \frac{dm}{dr}=4\pi r^2{\cal E}(r).
\end{equation}
By using the equation of state found in the previous section, we
integrate the TOV equations to calculate the structure of the
strange star \cite{n}. The results of this calculation are given
in the following figures and tables.

Figs. \ref{fig7} and \ref{fig8} show the gravitational mass versus
the central energy density at different values of temperature for
two cases of the constant and density dependent ${\cal B}$. For
each value of the temperature, these figures show that the
gravitational mass increases rapidly by increasing the energy
density and finally reaches to a limiting value at higher energy
densities. It is seen that the limiting value of the gravitational
mass increases by decreasing temperature.
Comparing Figs. \ref{fig7} and \ref{fig8}, one concludes that at
all temperatures, for the density dependent bag constant, the rate
of increasing mass with increasing the central density, at lower
values of the central densities, is substantially higher than that
of the case for fixed bag constant, especially at zero
temperature.
In Figs. \ref{figr90} and \ref{figrbn}, we have plotted the radius
of strange star versus the central energy density for both ${\cal
B}=90 \frac{MeV}{fm^3}$ and density dependent ${\cal B}$ at
different temperatures.
From Figs. \ref{fig7}$-$\ref{figrbn}, it can be seen that at each
central density, both mass and the corresponding radius increase
by decreasing the temperature.
The gravitational mass of strange star is also plotted as a
function of the radius for the constant and density dependent
${\cal B}$ in Figs. \ref{fig9} and \ref{fig10}. It is seen that
for all temperatures, the gravitational mass of strange star
increases by increasing the radius and it approaches a limiting
value (maximum mass).
Figs. \ref{fig9} and \ref{fig10} show that by decreasing the
temperature, the limiting values of mass and the corresponding
radius both increase.

In Tables \ref{tab1} and \ref{tab2}, the maximum mass and the
corresponding radius and central energy density of the strange
star at different temperatures for two cases of the constant and
density dependent ${\cal B}$ are given. It is shown that by
decreasing the temperature, the maximum mass of strange star
increases. This behavior is also seen for the radius of strange
star versus the temperature.
 Meanwhile, the central energy density decreases by decreasing the
temperature. By comparing Tables \ref{tab1} and \ref{tab2}, we can
see that for all temperatures, the maximum mass and the
corresponding radius calculated with the constant ${\cal B}$ are
less than those calculated with the density dependent ${\cal B}$.

\section{Summary and Conclusion}
We have considered a pure quark matter for the strange star to
calculate the structure properties of this object at finite
temperature. For this purpose, some thermodynamic properties of
the quark matter such as the entropy, free energy and the equation
of state have been computed using the constant and density
dependent bag constant (${\cal B}$). It was shown that the free
energy of the quark matter decreases by increasing the temperature
while the entropy of this system increases by increasing the
temperature. It was indicated that by increasing the temperature,
the equation of state of the quark matter becomes stiffer. We have
calculated the gravitational mass of the strange star by
numerically integrating the Tolman-Oppenheimer-Volkoff (TOV)
equations. Our results show that the gravitational mass of the
strange star increases by increasing the central energy density.
It was shown that this gravitational mass reaches a limiting value
(maximum mass) at higher values of the central energy density. We
have found that the maximum mass of the strange star decreases by
increasing the temperature. It was also shown that the maximum
mass and radius of the strange star in
 the case of density dependent ${\cal B}$ are higher than those in
 the case of constant ${\cal B}$.

\section*{Acknowledgements} {This work has been supported by Research
Institute for Astronomy and Astrophysics of Maragha. We wish to
thank Shiraz University Research Council.}


\newpage

\begin{table}
\begin{center}
\caption{Maximum mass ($M_{max}$) in solar mass unit
($M_{\odot}$), and the corresponding radius (R) and central energy
density (${\cal E}_{c}$) of the strange star at different
temperatures (T) for ${\cal B}=90\ \frac{MeV}{fm^3}$.}\label{tab1}

\begin{tabular}{cccc}
\hline
$T$ (MeV) & $M_{max}(M_{\odot})$ & R (km) &
${\cal E}_{c}(10^{14}\frac{gr}{cm^3})$ \\
  \hline
  0  & 1.354 & 7.698 & 38.24 \\
  30 & 1.228 & 7.073 & 47.54 \\
  70  & 1.101 & 6.416 & 60.60 \\
  80  & 1.039 & 6.142 & 63.65 \\
   \hline
\end{tabular}
\end{center}
\end{table}

\begin{table}
\begin{center}
\caption{As Table \ref{tab1} but for the density dependent ${\cal
B}$.}\label{tab2}

\begin{tabular}{cccc}
  \hline
  $T$ (MeV) & $M_{max}(M_{\odot})$ & R (km) &
  ${\cal E}_{c}(10^{14}\frac{gr}{cm^3})$ \\
  \hline
  0  & 1.676 & 8.761 & 39.11 \\
  30 & 1.341 & 7.442 & 48.47 \\
  70 & 1.181 & 6.768 & 61.56 \\
  80 & 1.122 & 6.567 & 64.21 \\
   \hline
\end{tabular}
\end{center}
\end{table}
\newpage

\begin{figure}
\includegraphics{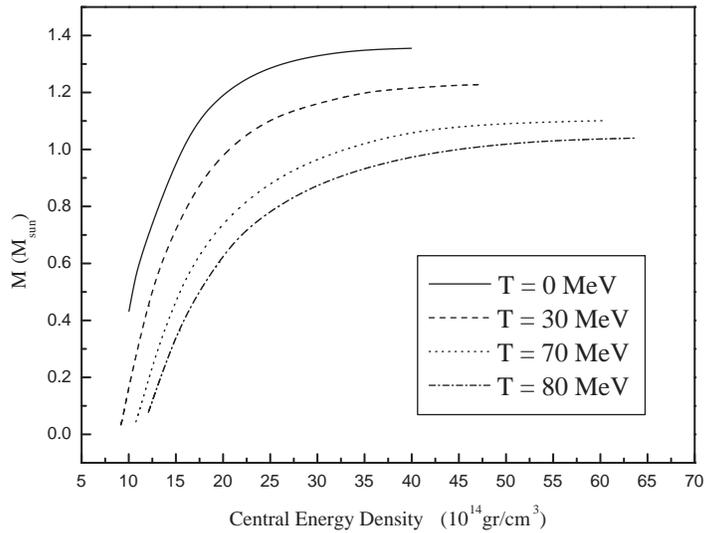}
\caption{The gravitational mass of the strange star as a function
of the central energy density at different temperatures for ${\cal
B}=90\ \frac{MeV}{fm^3}$.} \label{fig7}
\end{figure}

\begin{figure}
\includegraphics{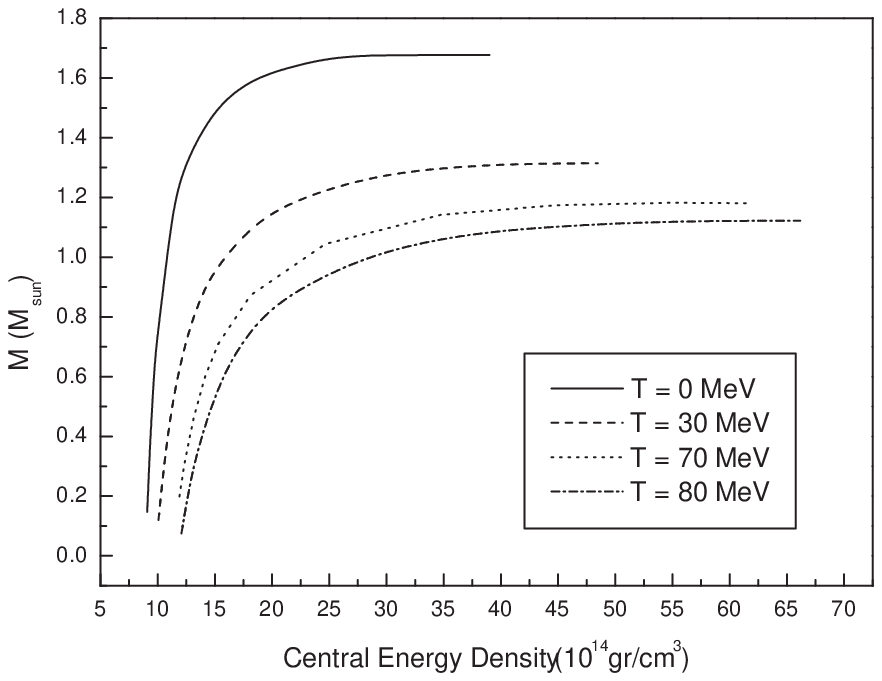}
\caption{As Figure \ref{fig7} but for the density dependent ${\cal
B}$. } \label{fig8}
\end{figure}

\begin{figure}
\includegraphics{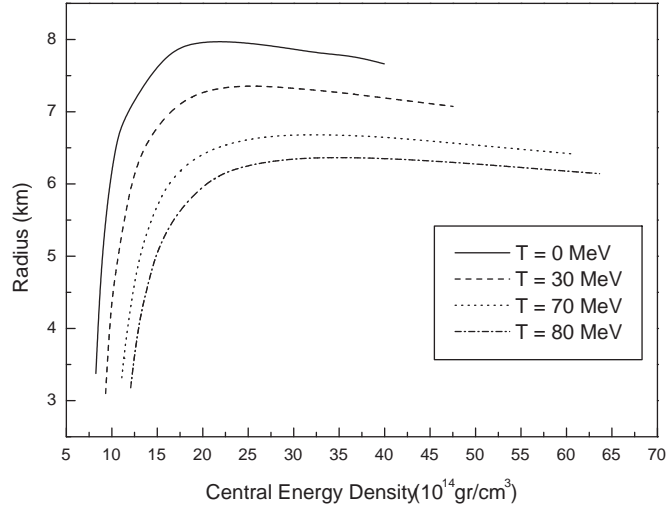}
\caption{The radius of the strange star as a function of the
central energy density at different temperatures for ${\cal B}=90\
 \frac{MeV}{fm^3}$.} \label{figr90}
\end{figure}

\begin{figure}
\includegraphics{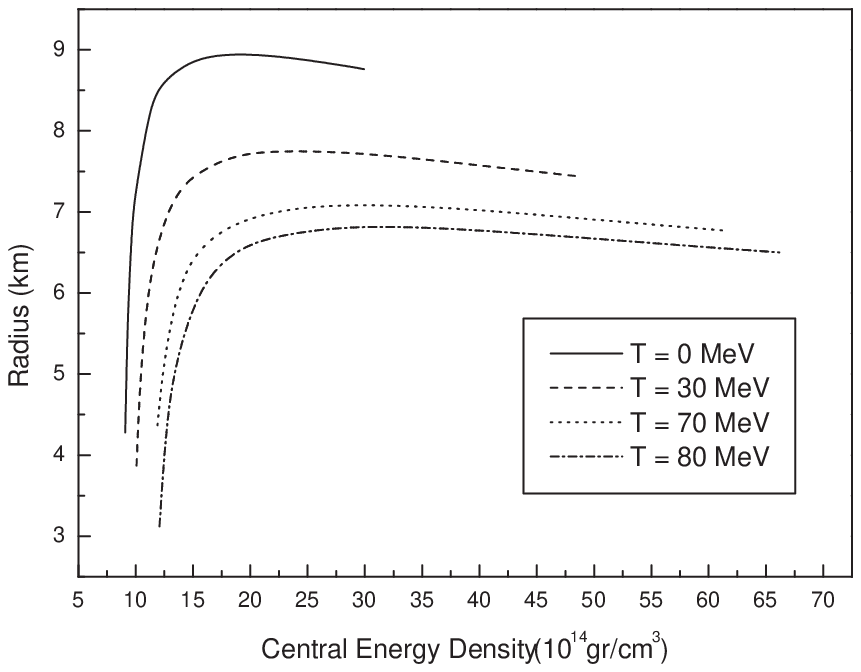}
\caption{As Figure \ref{figr90} but for the density dependent
${\cal B}$. } \label{figrbn}
\end{figure}

\begin{figure}
\includegraphics{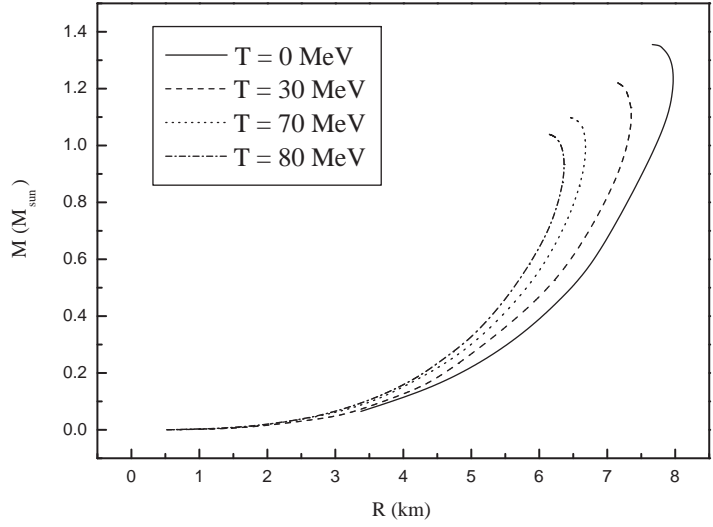}
\caption{The gravitational mass of the strange star as a function
of the radius at different temperatures for ${\cal B}=90\
\frac{MeV}{fm^3}$.} \label{fig9}
\end{figure}

\begin{figure}
\includegraphics{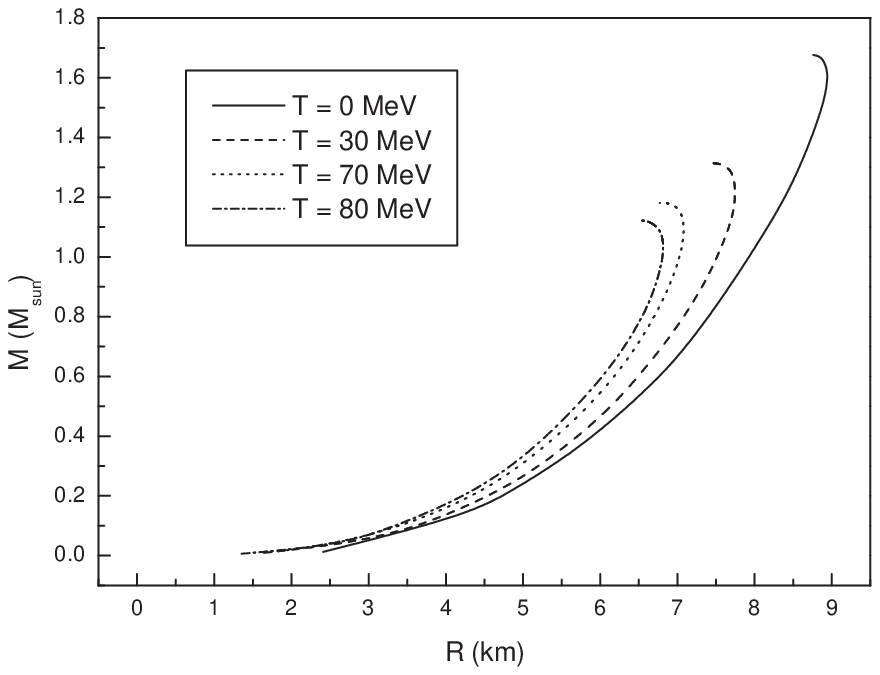}
\caption{As Figure \ref{fig9} but for the density dependent ${\cal
B}$.} \label{fig10}
\end{figure}

\end{document}